\documentclass[article,aps,twocolumn,showpacs,amsmath,amssymb,superscriptaddress,nofootinbib,floatfix]{revtex4-2}

\usepackage{graphicx}
\usepackage{dcolumn}
\usepackage{bm}
\usepackage[colorlinks=true,linkcolor=blue,citecolor=blue,urlcolor=blue]{hyperref}

\usepackage{array,booktabs,tabularx}
\usepackage[caption=false]{subfig}
\usepackage[USenglish]{babel}
\usepackage{braket}
\usepackage{xcolor}
\usepackage{amsfonts}
\usepackage{amssymb}
\usepackage[normalem]{ulem}
\usepackage{amsmath}
\usepackage[utf8]{inputenc}

\begin{document}

\title{Unravelling disorder in kagome Yb$_{0.5}$Co$_3$Ge$_3$}

\author{A. Korshunov}
\email{artem.korshunov91@gmail.com}
\affiliation{Donostia International Physics Center (DIPC), Paseo Manuel de Lardizábal. 20018, San Sebastián, Spain}

\author{C. A. Fuller}
\affiliation{Swiss-Norwegian BeamLines at European Synchrotron Radiation Facility, BP 220, F-38043 Grenoble Cedex, France}

\author{C.-Y. Lim}
\affiliation{Donostia International Physics Center (DIPC), Paseo Manuel de Lardizábal. 20018, San Sebastián, Spain}

\author{A. Kar}
\affiliation{Donostia International Physics Center (DIPC), Paseo Manuel de Lardizábal. 20018, San Sebastián, Spain}

\author{S. Roychowdhury}
\affiliation{Max Planck Institute for Chemical Physics of Solids, 01187 Dresden, Germany}
\affiliation{Department of Chemistry, Indian Institute of Science Education and Research Bhopal, Bhopal-462 066, India}

\author{D. Chernyshov}
\affiliation{Swiss-Norwegian BeamLines at European Synchrotron Radiation Facility, BP 220, F-38043 Grenoble Cedex, France}

\author{C. Shekhar}
\affiliation{Max Planck Institute for Chemical Physics of Solids, 01187 Dresden, Germany}

\author{A. Bosak}
\affiliation{European Synchrotron Radiation Facility (ESRF), BP 220, F-38043 Grenoble Cedex, France}

\author{C. Felser}
\affiliation{Max Planck Institute for Chemical Physics of Solids, 01187 Dresden, Germany}

\author{S. Blanco-Canosa}
\email{sblanco@dipc.org}
\affiliation{Donostia International Physics Center (DIPC), Paseo Manuel de Lardizábal. 20018, San Sebastián, Spain}
\affiliation{IKERBASQUE, Basque Foundation for Science, 48013 Bilbao, Spain}

\date{July 2025}

\begin{abstract}
The presence of phonon instabilities that leads to the formation of charge modulated states in kagome metals has allowed their classification based on the atomic displacements. Here, we use diffuse and inelastic x-ray scattering, backed by Monte Carlo simulations to describe a type-I instability in the kagome metal Yb$_{0.5}$Co$_3$Ge$_3$. We find that the in-plane distortion of Co in the kagome plane drives a structural transition with the appearance of new Bragg peaks at odd $L$, which are surrounded by a hexagonal diffuse signal. The anisotropic diffuse scattering, characteristic of a highly frustrated triangular lattice was simulated following a combination of Ising hamiltonian and Lennard-Jones potential, and demonstrate that the structural phase transition in Yb$_{0.5}$Co$_3$Ge$_3$ is of an order-disorder transformation type. The inelastic spectra reveals no softening but an anomalous broadening of the $\Gamma$–$A$ low energy acoustic mode. Our results highlight the critical role of the geometric frustration in promoting ordering from disorder in kagome lattices and power of diffuse scattering to disentangle the internal atomic displacements and correlated disorder.

\end{abstract}

\maketitle
\section{Introduction}

Disorder, whether it is intrinsic or extrinsic (defects, vacancies, stacking faults,...), is ubiquitous in solids. It is central between localized and metallic phases in the Anderson transition \cite{Evers_2008}, high Tc superconductivity in the cuprates \cite{Halloul_2009}, frustrated systems \cite{Grimm_1986} and the stabilization of metastable phases in \textit{glassy} solids \cite{Dasilva_2019,Uemura_1985}. In the case of intrinsic disorder, responsible for the formation of long range orders such as charge density wave (CDW) phases, it is divided into displacement (phonon driven) and occupational (substitutional) disorder \cite{Welberry_2004}. Displacement disorder is often found in displacive phase transitions, which are usually driven by a condensation of a soft mode \cite{Diego_2021,Weber_Tise2,Hoesch_2009} and have its roots mainly in nested segments of the Fermi surface \cite{Rossnagel_2011}, electron-phonon interaction \cite{Johannes_2008} or exciton formation \cite{Cercelier_2007}. On the other hand, the occupational disorder does not involve the freezing of a particular atomic displacement – although it does locally affect lattice vibrations – but rather the partial substitution of one or more atomic sites in the unit cell. Such occupational disorder, commonly observed in spin glasses \cite{Ozolins_1998,Okamoto_2016,Hultgren_1973}, ferroelectrics \cite{Glinchuk_2018,Yun_2022,Whitfield_2016}, high-T$_c$ cuprates \cite{Ackar_2014}, transition metal dichalcogenides \cite{Dyadkin_2015} and more recently in kagome metals \cite{subires2025frustrated}, present a challenge for the experimental and theoretical understanding of modulated states, and require a joint knit between experiment and theory.   

\begin{figure}
    \centering
    \includegraphics[width=\linewidth]{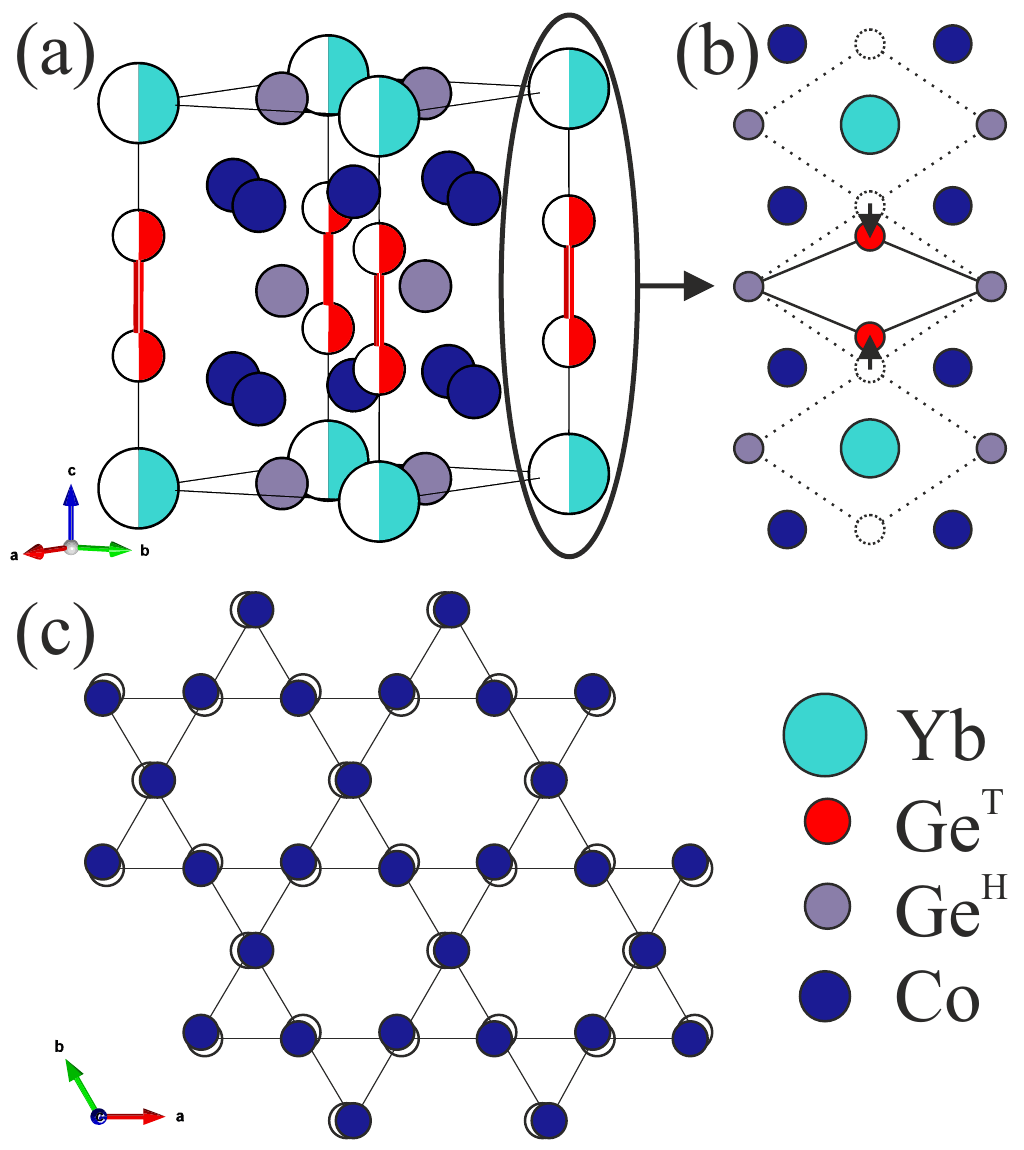}
    \caption{Crystal structure of Yb$_{0.5}$Co$_3$Ge$_3$. Blue spheres represent Co atoms, cyan spheres are Yb atoms, and red/grey spheres correspond to Ge atoms in trigonal/hexagonal states, respectively. (a) Average structure of the disordered HfFe$_6$Ge$_6$-type, with approximately 50\% occupancy at Yb and Ge$^\mathrm{T}$ sites. (b) Side view along the $c$-axis showing the Yb/Ge ordering and highlighting Ge dimerization. Dotted spheres denote atom positions in the ideal CoSn-type structure in the absence of inserted Yb atoms. (c) Top view of Yb$_{0.5}$Co$_3$Ge$_3$ highlighting the distortion of the Co kagome lattice resulting from the phase transition. Blue and white spheres represent Co atoms from adjacent layers.}
    \label{Fig1}
\end{figure}

Focusing particularly on the kagome lattice, this two-dimensional frustrated network of corner-sharing triangles emerges as an ideal platform to study the impact of disorder in its ground state \cite{Nakatsuji_2005,Shokef_2011}, and a fertile platform for exploring novel quantum phenomena \cite{Wang_2023}. Its electronic band structure, featuring flat bands, Dirac band crossings and van Hove singularities, gives rise to strongly correlated electron states and topologically nontrivial excitations 
\cite{negi2025magnetic,neupert2022charge,kang2022twofold,guguchia2023hidden}. Depending on the electron filling, a variety of emergent phases have been discovered, including magnetic order, charge-density waves, nematic order, superconductivity and associated topological states \cite{Kiesel_2013,Denner_2022,Wu_2023,neupert2022charge,guo2009topological}. Moreover, geometric frustration can suppress conventional magnetic order and even promote quantum spin-liquid behavior \cite{balents2010spin,zhou2017quantum}. The delicate interplay of lattice, spin and charge degrees of freedom makes kagome materials an exciting platform to study correlation-driven physics and the impact of intrinsic disorder in their ground state. 

An example lies in the family of AV$_3$Sb$_5$ kagome metals (A = K, Rb, Cs) that host multiple-q CDW connecting the M and L points of the Brillouin zone (BZ) \cite{Ortiz_2019,Ortiz_2022}, mirroring the imaginary phonon modes predicted by Density Functional Theory (DFT) \cite{Subedi_2022}. The controversial phase transition has been described as unconventional \cite{Miao_2021}, order-disorder  \cite{Subires_2023} and anharmonic scenarios \cite{Gutierrez_2024,Liu_2022,He_2024}, having in common the condensation of a transverse vibration of the V atoms in the kagome plane. Moreover, the introduction of chemical disorder by substituting Ti or Sn in the kagome net, besides modifying the carrier concentration, leads to a suppression of the interlayer charge correlations and the CDW order \cite{Huai_2024}, and the formation of incommensurate, quasi-1D charge correlations \cite{Kautzsch_2023}. 

Going beyond the 135 series, the 166 family (MT$_6$Z$_6$-type) hosts three different types of CDWs depending on the particular atomic vibrations involved in the CDW transition \cite{Feng_2024}. Type-I CDWs are characterized by an in-plane distortion of kagome net, while type-II and type-III feature in-phase and out-of-phase out-of-plane distortion of trigonal M and Z atoms, respectively. Type-III CDWs have identified in 166 ScV$_6$Sn$_6$ \cite{Arachchige_2002} and related compounds \cite{Ortiz_2025}, and in FeGe \cite{Teng_2022,Teng_2023}. Microscopically, their lattice dynamics show signatures of a phonon softening of such longitudinal out-of-plane modes \cite{lee2024nature,Cao_2023,hu2025flat,hu2024phonon,korshunov2023softening} and occupational disorder scenarios \cite{alvarado2024frustrated}. Type-I CDWs are predicted to be present in the 166-(Ni,Co)Ge \cite{wang2022electronic,weiland2020refine} and 166-NiSi, but the nature of the phase transition and its dynamics, contained in the high temperature charge fluctuations that preempt the CDW phase transition, have not been addressed yet. This is, for instance, exemplified in the disordered Yb$_{0.5}$Co$_3$Ge$_3$, whose lattice dynamics show imaginary modes along the $\Gamma$-A line that describe in-plane out-of-phase displacements of the kagome sites \cite{Feng_2024}. Indeed, DFT calculations reveal an enhanced density of states (DOS) as a result of the presence of higher order saddle points in the electronic structure and a \textit{flattening} the band dispersion \cite{Tan_2024}. 

In this work, we focus on Yb$_{0.5}$Co$_3$Ge$_3$, that has been predicted to host type-I phonon instabilities \cite{Feng_2024}, and study the microscopic origin of the structural phase transition. We use a combination of diffuse scattering (DS), Monte Carlo (MC) simulations, and inelastic x-ray scattering (IXS) to unveil a structural phase transition, driven by the in-plane displacement of the Co atoms that alternately twist the kagome plane. On the other hand, the partial substitution of the Yb and Ge atoms appears as anisotropic hexagonal diffuse rings in the diffraction pattern, a hallmark of strong competition between inclusion atoms in interlayer void sites, caused by frustration inherent in the triangular packing. This anisotropic scattering is well captured by a modified Ising model, describing an anisotropic in-plane ordering of Yb and trigonal Ge (Ge$^\mathrm{T}$) atoms. Furthermore, the structural transition does not involve any phonon softening, but rather an anomalous broadening of the $\Gamma$–A phonon branch. Our results indicate that there are two independent processes that dictate the ground state in Yb$_{0.5}$Co$_3$Ge$_3$: a structural transition originating from the distortion of the Co kagome lattice and an ordering of Yb atoms and Ge dimers with subsequent distortion of neighboring hexagonal Ge (Ge$^\mathrm{H}$) atoms in the \textit{ab}-plane. On more general grounds, we highlight the ability of diffuse scattering to unravel the coherent atomic displacements in correlated disordered lattices and the deviations from a perfect crystal.

\begin{figure*}
    \centering
    \includegraphics[width=1.0\linewidth]{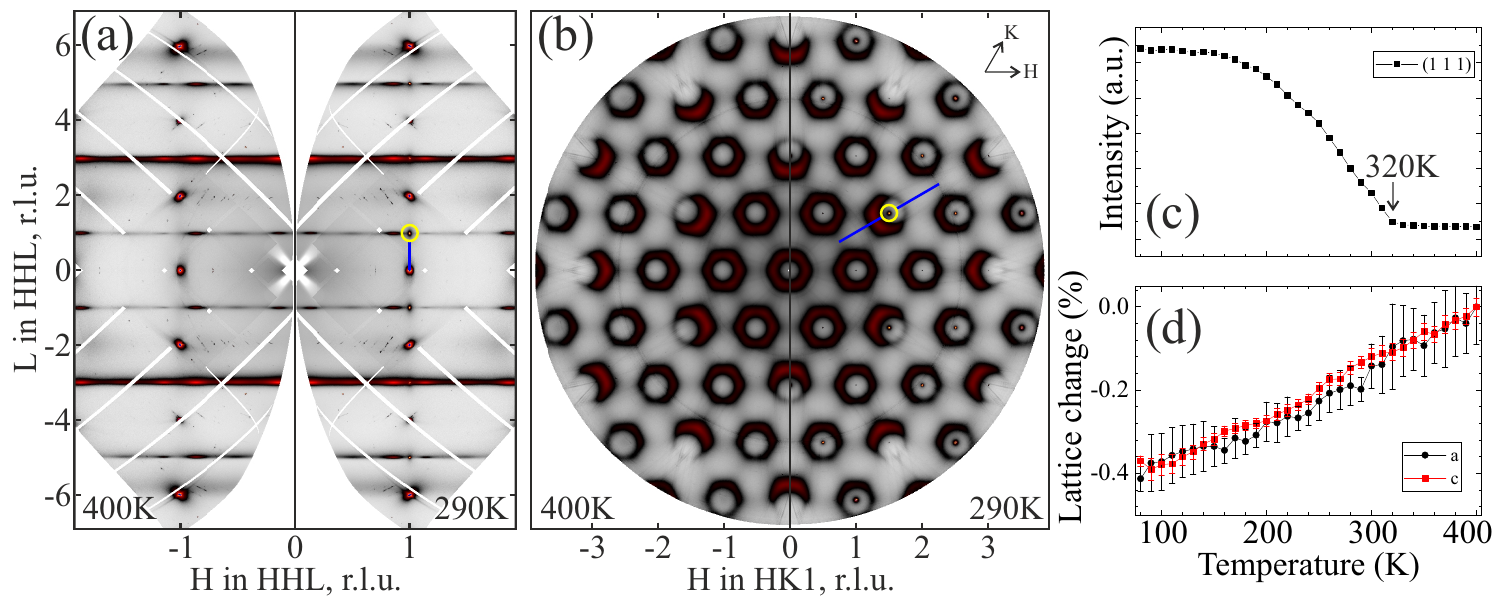}
    \caption{Diffuse scattering maps of Yb$_{0.5}$Co$_3$Ge$_3$ in the $HHL$ (a) and \textit{HK}1 (b) planes, measured at 290~K and 400~K. Appropriate Laue symmetry was applied to the reconstructed layers to remove the detector gaps and improve signal-to-noise ratio. The yellow circle highlights the (1~1~1) satellite, and the blue lines indicate the 11\textit{L} out-of-plane (a) and \textit{HH}1 in-plane directions used for IXS measurements. (c) Temperature evolution of the DS integrated intensity around the (1~1~1) reflection. (d) Temperature dependence of the $a$- and $c$-axis lattice parameters.}
    \label{Fig2}
\end{figure*}

\section{Methods}

Needle-like single crystals of Yb$_{0.5}$Co$_3$Ge$_3$ were grown using the flux method, as reported elsewhere \cite{weiland2020refine}. Temperature dependent DS measurements were carried out at the ID28 beamline of the European Synchrotron Radiation Facility (ESRF), using a wavelength of 0.6968~\AA\ (17.8~keV). Diffraction patterns were collected with a Dectris PILATUS3 1M area detector. The temperature was controlled using a Cryostream 700+ N$_2$ gas-flow cooler (Oxford Cryosystems). The orientation matrix was refined using CrysAlisPro software, and the beamline-specific software ProjectN was used for high-resolution reciprocal space map reconstructions. IXS measurements were performed on the same crystal at the IXS branch of the ID28 beamline. The sample temperature was controlled using the same Cryostream device as in the diffraction experiment. The spectrometer operated with a Si(999) backscattering monochromator at a wavelength of 0.6968~\AA\ (17.8~keV), providing an energy resolution of 3~meV. IXS energy-transfer scans (at constant momentum transfer) were obtained in transmission geometry along selected high-symmetry directions of reciprocal space. The components $(H\ K\ L)$ of the scattering vector are expressed in reciprocal lattice units (r.l.u.): $(H\ K\ L) = H\mathbf{a}^* + K\mathbf{b}^* + L\mathbf{c}^*$, where $\mathbf{a}^*$, $\mathbf{b}^*$, and $\mathbf{c}^*$ are the reciprocal lattice vectors.

All IXS spectra were fitted using a single damped harmonic oscillator model. The spectral components (elastic line and phonon excitations) were extracted by deconvoluting the spectra with the experimental resolution function. The energy position, linewidth, and intensity of each peak were used as fitting parameters. The Stokes and anti-Stokes intensities were corrected for the Bose–Einstein thermal population factor. 
MC simulations were performed to generate real-space realizations of atomic configurations that could reproduce the observed diffuse scattering. The average Yb$_{0.5}$Co$_3$Ge$_3$ unit cell was expanded into a $50 \times 50 \times 50$ supercell, with 50\% of the 1\textit{a} sites randomly occupied by Ge dimers (represented as Ising variables $\sigma = +1$) and 50\% by Yb atoms ($\sigma = -1$). In each MC move, two Ising variable were inverted, and the move was accepted or rejected according to the Metropolis algorithm. MC simulations were started from different randomized configurations to minimize the risk of becoming trapped in local energy minima. Each simulation was run until the energy converged, and the resulting atomic configuration was used to calculate the average diffuse scattering in the DISCUS program~\cite{Proffen:hw0052}.

\section{Results}

Yb$_{0.5}$Co$_3$Ge$_3$ adopts a disordered HfFe$_6$Ge$_6$-type structure (space group $P6/mmm$) at room temperature, as shown in Fig.~\ref{Fig1}(a), where the Co atoms form ideal kagome sheets and the Yb and Ge atoms occupy the interlayer voids in a disordered manner \cite{dzyanyj1995crystal,weiland2020refine,wang2022electronic,meduri2025evolution}. The compound was found to be paramagnetic, without any trace of magnetic order down to 2 K \cite{weiland2020refine}, however, a structural transition sets in at low temperature, as a result of the alternate twisting of the Co triangles within the kagome plane, Fig.~\ref{Fig1}(c), such that the equilateral Co–Co bonds become inequivalent \cite{wang2022electronic}. This deformation breaks the sixfold rotational symmetry and the mirror planes, which results in lowering crystal symmetry from $P6/mmm$ (No.~191) to $P6_3/m$ (No.~176) space group. Notably, the Yb and Ge sublattices remain disordered and largely retain their high-temperature average positions, suggesting that the transition primarily involves the Co kagome layer and the connectivity between layers \cite{weiland2020refine}.

\begin{figure*}
    \centering
    \includegraphics[width=\linewidth]{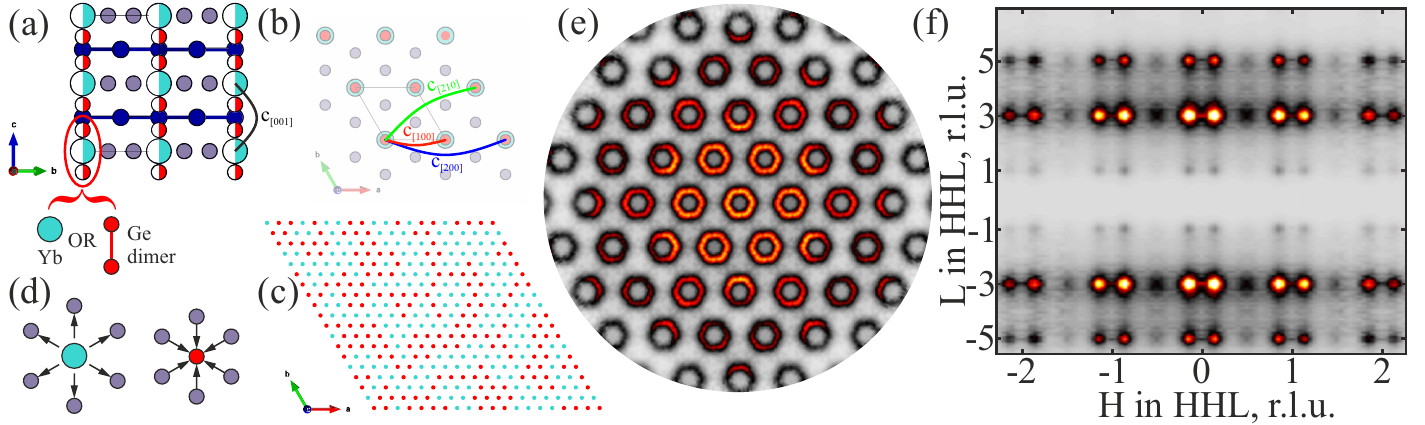}
    \caption{(a) Sketch of the model used in the MC simulations. Out-of-plane and in-plane correlations are illustrated in panels (a) and (b), respectively. (c) Simulated distribution of Yb atoms and Ge-Ge dimers from the MC results. (d) Local in-plane atomic distortions. Calculated DS from MC simulation in \textit{HK}1 (e) and \textit{HHL} map (f).}
    \label{Fig3}
\end{figure*}


Fig.~\ref{Fig2}(a-b) summarizes the DS maps for Yb$_{0.5}$Co$_3$Ge$_3$ in the $HHL$ and $HK1$ reciprocal space planes, measured at 290~K and 400~K. The measurements were performed with high incident flux, which led to oversaturation of the Bragg reflections, but enhanced the visibility of weak DS features. Sharp Bragg peaks corresponding to the main phase are clearly visible at even $L$-planes ($L = 0, 2, 4\dots$), consistent with the average crystal structure described by the hexagonal space group $P6/mmm$, with lattice parameters $a \approx 5.1$~\AA{} and $c \approx 7.8$~\AA. At high temperature, the DS pattern reflects a disordered YCo$_6$Ge$_6$-type structure \cite{weiland2020refine} without long-range ordering of Yb and Ge atoms along the $c$-axis. Upon cooling below T$^*$ = 320~K, a structural phase transition occurs, involving a distortion of the Co-based kagome network and the emergence of superstructure reflections in odd $L$-planes ($L = 1, 3, 5\dots$), highlighted in yellow circles in Fig.~\ref{Fig2}(a-b), consistent with previous findings \cite{wang2022electronic}. The intensities of the lower symmetry superstructure satellites increase upon cooling in a continuous fashion, indicating the gradual development of long-range structural order, Fig.~\ref{Fig2}(c). The low-temperature phase is described within the $P6_3/m$ space group, characterized by a distorted kagome lattice with the doubling of the \textit{c}-axis. No significant anomalies were detected in the temperature dependence of the lattice parameters, aside from the expected thermal contraction upon cooling, as sketched in Fig.~\ref{Fig2}(d).

\begin{figure*}
    \centering
    \includegraphics[width=1.0\linewidth]{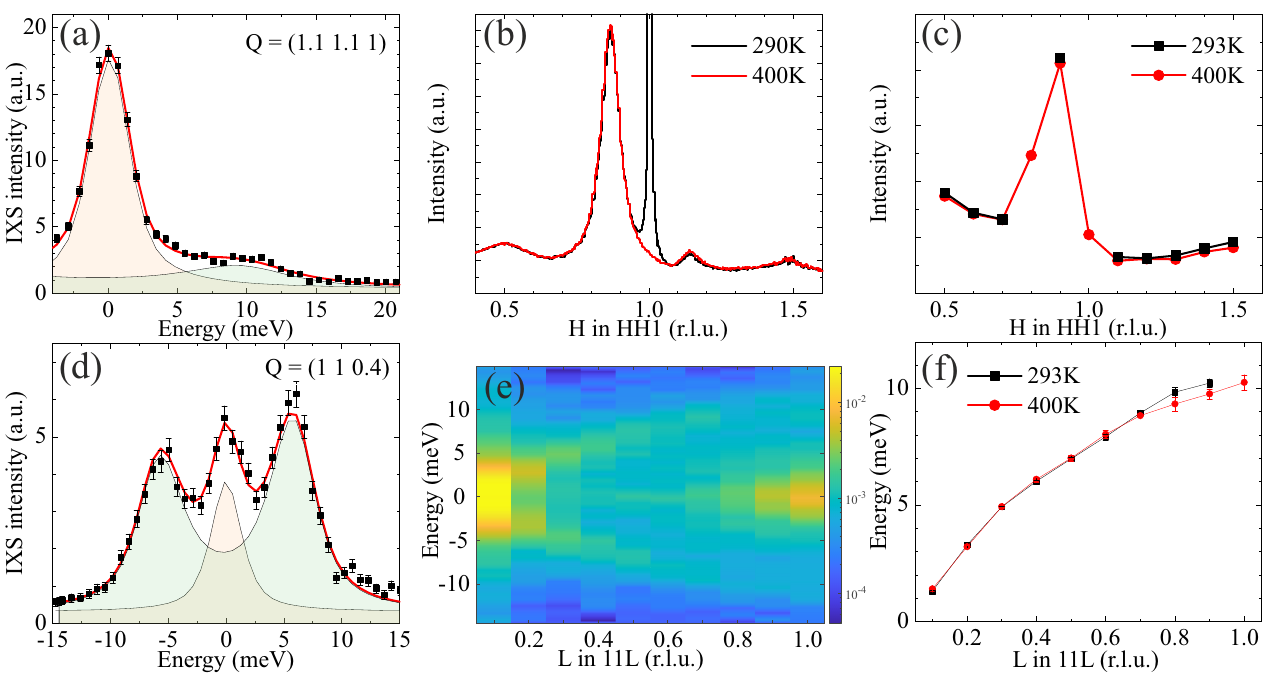}
    \caption{Lattice dynamics of Yb$_{0.5}$Co$_3$Ge$_3$. (a) In-plane IXS spectrum with a strong elastic peak and a broad $\Gamma$–$K$ optic phonon. Black dots with error bars in (a) and (d) represent experimental data; the red solid lines are fits, modeled as convolutions with the experimental resolution $\Delta E \approx 3$~meV. Shaded areas indicate individual contributions to the spectra. (b) DS intensity profile along the [$HH1$] direction (blue line in Fig.~\ref{Fig2}(b)), with a prominent (1~1~1) peak arising from the Co kagome lattice ordering. (c) Momentum dependence of the elastic line intensity extracted from in-plane IXS spectra, confirming the elastic nature of the DS. The peak corresponds to the DS maximum observed in (b). (d) Out-of-plane IXS spectrum showing the elastic line (zero energy loss), and Stokes and anti-Stokes components of the $\Gamma$–$A$ acoustic phonon. (e) Energy–momentum IXS map along the (1 1 \textit{L}) direction at 400~K, highlighting acoustic phonon dispersion and increase of elastic line intensity in proximity of the diffuse plane.  (f) Momentum dependence of the (1~1~\textit{L}) phonon energy at RT and 400~K.}
    \label{Fig4}
\end{figure*}

Further investigation of the DS reveals a strong anisotropic intensity forming hexagonal clouds with a radius of approximately $a^*/4$, a hallmark of frustrated triangular lattices \cite{Tosic_2024,subires2025frustrated}. The diffuse signal is observed at all odd $L$ planes surrounding the emerging superstructure peaks and remains temperature-independent. Additionally, the DS intensity decreases radially in $HK$ planes (see Fig.~\ref{Fig2}(b)), which is a clear signature of substitutional disorder-induced short-range correlations \cite{subires2025frustrated,Tosic_2024}. The correlation lengths derived from the width of the diffuse scattering are approximately 30~\r{A} in-plane and 140~\r{A} along the out-of-plane direction. As indicated previously, this DS arises from the stochastic packing of Yb atoms and Ge–Ge dimers along the $c$-direction, resulting in a doubling of $c$-axis, see also reference \cite{weiland2020refine}. Notably, the intensity of DS in the $HK1$ plane is substantially lower than in $HK3$, resembling the pattern expected for HfFe$_6$Ge$_6$-type ordering. In this structural motif, the structure factor for reflections $(HKL)$ reaches a maximum at $L = 3n$, with intensity roughly proportional to the sum of atomic numbers, $\left( Z_{\text{Yb}} + 2Z_{\text{Ge}} \right)$ \cite{venturini2006filling}. Such short-range modulations cannot be captured by conventional structural refinements based solely on Bragg intensities. In contrast, DS provides rich information on both in-plane and out-of-plane correlations, revealing subtle structural distortions beyond the average crystallographic picture. 

\begin{figure}
    \centering
    \includegraphics[width=0.9\linewidth]{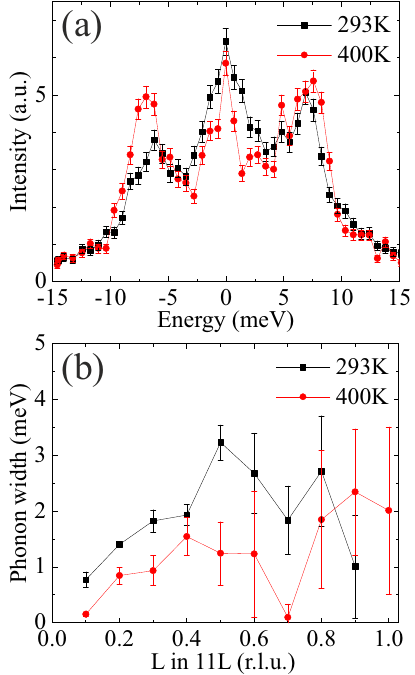}
    \caption{(a) Comparison of IXS spectra at the (1~1~0.5) position, highlighting temperature-induced phonon broadening. (b) Momentum dependence of the (11\textit{L}) phonon linewidth extracted from fits at RT and 400~K.}
    \label{Fig5}
\end{figure}

To gain insight into the plausible three-dimensional ordering within the crystal and microscopic origin of the anisotropic DS, we performed MC simulations incorporating a binary substitutional disorder to order the Yb/Ge atoms, Fig.~\ref{Fig3}(a). This disorder can be effectively described by an Ising model on a triangular antiferromagnetic lattice \cite{subires2025frustrated,korshunov2025cascade,Fuller_2022,Welberry_2007}. The corresponding Hamiltonian is given by:

\begin{equation}
    \begin{split}
    H = \sum_{<i,j>}c_{[100]}\sigma_i\sigma_j + \sum_{<i,j>}c_{[210]}\sigma_i\sigma_j +   \sum_{<i,j>}c_{[200]}\sigma_i\sigma_j + \\
    \sum_{<i,j>}c_{[001]}\sigma_i\sigma_j,
    \end{split}
\label{eq:hamiltonian}
\end{equation}


where the Ising variables $\sigma = \pm1$ stand for the occupation of a structural void by either a Yb atom ($+1$) or a Ge–Ge dimer ($-1$). The initial configuration assumes a 50\% random distribution of these atoms on the (0 0 0) site, which was fixed during the simulation, consistent with the site occupancy extracted from the average crystallographic unit cell. The strength and direction of interactions are controlled by the coefficients $c_{[abc]}$ of the Ising \textit{pseudo}-spins, initially assuming correlations with four nearest neighbors, Fig.~\ref{Fig3}(a-b).
After exploring the parameter space, we found that a combination of correlations with the first in-plane neighbors ($c_{[100]} = -0.4$ meV, $c_{[210]} = 0.4$ meV, and $c_{[200]} = 0.133$ meV), together with interlayer correlation, $c_{[001]} = 0.5$ meV, best reproduces the experimental diffuse scattering. These values reflect the frustrated nature of the trigonal ordering, characterized by a competition between positive and negative in-plane correlations and a dominant antiferromagnetic-like coupling between adjacent layers along the \( c \)-direction. A representative atomic configuration obtained from the simulation is shown in Fig.~\ref{Fig3}(c), exhibiting a stripe-like pattern. Within the $ab$ plane Ge dimers tend to form small clusters/channels spanning around 2-4 unit cells. Along the $c$-axis, there is a tendency to order in an alternating pattern, where Ge dimers prefer to be adjacent to Yb rather than another dimer.

Simulated DS for the $HK1$ and $HHL$ planes successfully reproduce the characteristic hexagonal arrangement of diffuse clouds with its intensity falling as a function of momentum, although being isotropic in the azimuthal angle. To account for the in-plane anisotropy observed in the experimental DS, we also included relaxation of hexagonal Ge$^\mathrm{H}$ atoms at the \textit{2c} Wyckoff site ($\frac{1}{3}$\ $\frac{2}{3}$\ 0). These Ge$^\mathrm{H}$ atoms tend to shift towards the Ge dimers and away from the Yb atoms, due to the size-effect correlations, Fig.~\ref{Fig3}(d). This distortion was modeled via a Lennard-Jones potential:

\begin{equation}
E = \frac{A}{(r + \delta)^{12}} - \frac{B}{(r + \delta)^6},
\end{equation}

where $r$ is the equilibrium interatomic distance derived from the average structure, and $\delta$ is an empirical displacement parameter. The values $A$ and $B$ were chosen to reproduce a potential minimum at a distance $r + \delta$. Inclusion of this small in-plane distortion enabled the modeling of in-plane suppression of DS intensity, when moving outward from the $L$-axis, as shown in Fig.~\ref{Fig3}(e). Moreover, the distribution of DS intensity in the $HHL$ map is consistent with the experimentally observed one plotted in Fig.~\ref{Fig2}(a). Thus, our simulations support a picture in which Yb/Ge site disorder on the 1a sublattice exhibits frustrated short-range order, and a slight displacement of neighboring atoms is required to fully capture the observed DS intensity distribution.

We now examine the nature of the superstructure reflections that arise from the formation of a superlattice with a doubled unit cell along the \textit{c}-axis. To disentangle the static and dynamic contributions to the total scattering, we performed IXS measurements in the vicinity of the $(1\,1\,1)$ Bragg peak, as indicated by the blue line in Fig.~\ref{Fig2}(a-b). The in-plane IXS spectra consist of a strong quasi-elastic central peak (E$_\mathrm{loss}$=0) and a weak low-energy optic phonon, Fig.~\ref{Fig4}(a). Similarly, the diffuse plane reveals itself as a significant increase in the elastic line intensity upon entering the $L$ = 1 Bragg reflection. The intensity of the central peak closely follows the momentum dependence of the integrated hexagonal DS cloud observed in the $HK1$ reciprocal space map, Fig.~\ref{Fig4}(b) and (c). These observations indicate that the structural transition is essentially elastic in nature and originates from short-range structural ordering with a finite correlation length, due to the additional ordering along the \textit{c}-direction. On the other hand, the out-of-plane acoustic phonon, shown in Fig.~\ref{Fig4}(d), has higher intensity but does not exhibit any anomaly in its dispersion, Fig.~\ref{Fig4}(e) and (f). Furthermore, both in-plane and out-of-plane phonon frequencies do not follow the anharmonic behavior and remain virtually unchanged over the measured temperature range, but agree with theoretical phonon dispersions \cite{Feng_2024}. However, our IXS data above and below the structural transition, Fig.~\ref{Fig5}(a-b) show an increase of the out-of-plane phonon linewidth in a broad range of momentum along the $L$ direction, presumably associated to the in plane displacement of Co in the kagome plane.

\section{Discussion and Conclusions}

We have shown that disordered kagome Yb$_{0.5}$Co$_3$Ge$_3$ is characterized by a hexagonal diffuse scattering in the $HK$ planes with odd $L$, where the hexagons are centered by sharp superstructural Bragg spots, indicating a structural ordering with doubling of the unit cell. This ordering originates from the complex packing of Yb atoms and Ge dimers, which form alternating rods along the $c$-axis with pronounced disorder in the \textit{ab}-plane. Individual Yb/Ge rods are embedded in a honeycomb host lattice of hexagonal Ge$^\mathrm{H}$ atoms, forming a triangular packing pattern reminiscent of a frustrated triangular antiferromagnet with Ising-type interactions. The eventual selection of the ground state of the kagome net will be dictated by Fermi surface instabilities or the subtle wavevector-dependent electron-phonon interaction (EPI). In the absence of strong coupling or significant electronic correlations, the system could undergo an \textit{order-by-disorder}  mechanism \cite{subires2025frustrated,alvarado2024frustrated} or remain disordered at low temperatures. Our findings provide an experimental complement to recent theoretical work by Tan et al. \cite{Tan_2024}, who identified that substitutional disorder at the Yb site can directly influence the stability and observability of the $\zeta_1$ HOSP-related flatbands. This may reduce the electronic localization and correlation effects typically associated with kagome flatband physics. In contrast, phonon modes remain relatively unaffected, suggesting that the disorder alters electronic properties without inducing vibrational instabilities. Consistently, no clear phonon anomaly is observed in our IXS data. Instead, we detect anomalous broadening of out-of-plane modes, consistent with disorder-induced anharmonicity. This behavior indicates that the transition is not driven by a conventional soft-mode mechanism, but rather reflects an order–disorder scenario associated with short-range occupational disorder and frustrated local displacements. Such transitions likely involve shallow or anharmonic potential energy surfaces that broaden phonon linewidths without reducing phonon energies. Moreover, we attribute the higher observed T$^*$ compared to that reported in ~\cite{wang2022electronic} to less disorder in our sample, highlighting that disorder indeed suppresses the transition temperature, as observed in related kagome compounds \cite{PhysRevB.107.125124, PhysRevLett.132.256501}.

In summary, Yb$_{0.5}$Co$_3$Ge$_3$ provides a compelling case study of how structural distortions in a kagome metal – encompassing Co lattice distortions, potential Ge dimerization or ordering, and Yb/Ge site ordering into an alternating chain – can substantially alter the electronic ground state. Our comprehensive elastic and inelastic scattering results reveal that both occupational and displacement disorder play key roles in the observed structural transformation. This seems to be rather general in triangular lattices, such as the 166 ScV$_6$Sn$_6$ \cite{alvarado2024frustrated}, LuNb$_6$Sn$_6$ \cite{Ortiz_2025,Yang_2025}, and is expected to become even more important in kagome materials consisting on late transition metals (Ni,Cu), where steric effects may dominate the spectrum of phase transitions. More generally, we emphasize the key role of diffuse scattering as a probe of local correlations in crystalline materials, providing information on how different atoms coherently interact beyond the average structure.

\section{Acknowledgments}
D.S., A.Kar and S.BC. acknowledge financial support from the MINECO of Spain through the project PID2021-122609NB-C21 and by the European Union Next Generation EU/PRTR-C17.I1, as well as by IKUR Strategy under the collaboration agreement between Ikerbasque Foundation and DIPC on behalf of the Department of Education of the Basque Government. A.K. thanks the Basque government for financial support through the project PIBA-2023-1-0051. C.-Y.L. is supported by the European Research Council (ERC) under the European Union’s Horizon 2020 research and innovation program (grant agreement no. 101020833). C.F. acknowledges support from the DFG under SFB 1143 (Project No. 247310070), the Würzburg-Dresden Cluster of Excellence on Complexity and Topology in Quantum Matter - ct.qmat (EXC 2147, Project No. 390858490), and FOR 5249 (QUAST, Project No. 449872909). 

\bibliography{main_references}

\end{document}